\documentstyle[12pt]{article}
\newlength{\extraspace}
\newlength{\extraspaces}
\newcommand{\ba}{\begin{eqnarray}
\addtolength{\abovedisplayskip}{\extraspaces}
\addtolength{\belowdisplayskip}{\extraspaces}
\addtolength{\abovedisplayshortskip}{\extraspace}
\addtolength{\belowdisplayshortskip}{\extraspace}}
\newcommand{\ea}{\end{eqnarray}}

\newcommand{\nonu}{\nonumber \\[.5mm]}
\newcommand{\A}{&\!\!\!}

\begin{document}

\thispagestyle{empty}

\hfill \parbox{3.5cm}{hep-th/0211187 \\ SIT-LP-02/11 \\ (revised)}
\vspace*{1cm}
\vspace*{1cm}
\begin{center}
{\bf  ON LINEARIZATION OF SUPERON-GRAVITON MODEL(SGM)} \\[20mm]
{Kazunari SHIMA and Motomu TSUDA} \\[2mm]
{\em Laboratory of Physics, Saitama Institute of Technology}
\footnote{e-mail:shima@sit.ac.jp, tsuda@sit.ac.jp}\\
{\em Okabe-machi, Saitama 369-0293, Japan}\\[2mm]
{Manabu SAWAGUCHI} \\[2mm]
{\em High-Tech Research Center, Saitama Institute of Technology}
\footnote{e-mail:sawa@sit.ac.jp}\\
{\em Okabe-machi, Saitama 369-0293, Japan}\\[2mm]
{May 2003}\\[15mm]


\begin{abstract}
Referring to the supermultiplet of N=1 supergravity(SUGRA) 
the linearization of N=1 SGM action describing the nonlinear 
supersymmetric(NL SUSY) gravitational 
interaction of superon(Nambu-Goldstone(N-G) fermion) is attempted. 
The field contents of on-shell SUGRA supermultiplet are realized 
as the composites, though they have new 
SUSY transformations which closes on super-Poincar\'e(SP) algebra. 
Particular attentions are paid to the local Lorentz invariance. 

\end{abstract}
\end{center}

\newpage
Extending the geometrical arguments of Einstein general relativity theory(EGRT) on Riemann spacetime 
to new spacetime where the coset space coordinates of ${superGL(4,R) \over GL(4,R)}$ 
turning to the N-G fermion degrees of freedom(d.o.f.) are attached at every Riemann spacetime point,  
we have proposed a new Einstein-Hilbert(E-H) type action\cite{ks1}.  
The new E-H type action describes the NLSUSY\cite{va} invariant gravitational interaction of N-G fermion 
$superon$ in Riemann spacetime.   \par
%
%
%
%
%
%
In this letter we would like to discuss the linearization of the new E-H type action($N=1$ SGM action) 
to obtain the equivalent linear(L) SUSY\cite{wzgl} theory in the low energy, which is renormalizable. \\
Considering a phenomenological potential of SGM, though qualitative and group theoretical, 
discussed in \cite{ks2} based upon the composite picture of L SUSY representation 
and the recent interest in NLSUSY in superstring(membrane) world, 
the linearization of NLSUSY in curved spacetime may be of some general interest.  \par 
The linearization of SGM is physically interesting in general, even if the consequent theory 
were a existing SUGRA-like, for the equivalence among them gives a new insight into 
the fundamental structure of nature.
For the  self-contained arguments we review SGM action briefly.
($N = 1$) SGM action is given by\cite{ks1}; 
\begin{equation}
L_{SGM}=-{c^{3} \over 16{\pi}G}\vert w \vert(\Omega + \Lambda),
\label{SGM}
\end{equation}
\begin{equation}
\vert w \vert = {\rm det}{w^{a}}_{\mu} 
= {\rm det}({e^{a}}_{\mu}+ {t^{a}}_{\mu}),  \quad
{t^{a}}_{\mu} = {i \over 2} \kappa^4 (\bar{\psi}\gamma^{a}
\partial_{\mu}{\psi}
- \partial_{\mu}{\bar{\psi}}\gamma^{a}{\psi}),
\label{w}
\end{equation} 
where $e^{a}{_\mu}$ is the  vierbein of EGRT, 
$\psi$ is N-G fermion(superon), 
${\kappa^{4} = ({c^{3}\Lambda \over 16{\pi}G}})^{-1} $ 
is the fundamental volume of four dimensional spacetime 
of Volkov-Akulov(V-A) model\cite{va}, 
and $\Lambda$ is the ${small}$ cosmological constant 
related to the strength of the superon-vacuum coupling constant. 
$\Omega$ is a new scalar curvature analogous 
to the Ricci scalar curvature $R$ of EGRT, 
whose explicit expression is obtained  by just replacing 
${e^{a}}_{\mu}(x)$  by ${w^{a}}_{\mu}(x)$ in Ricci scalar $R$. 
These results can be understood intuitively by observing that 
${w^{a}}_{\mu}(x) ={e^{a}}_{\mu}(x)+ {t^{a}}_{\mu}(x)$  defined by 
$\omega^{a}={w^{a}}_{\mu}dx^{\mu}$, where $\omega^{a}$ is the NL SUSY 
invariant differential one-form of V-A\cite{va}, is invertible and 
$s^{\mu \nu}(x) \equiv {w_{a}}^{\mu}(x) w^{{a}{\nu}}(x)$ 
are a unified vierbein and a unified metric tensor 
in SGM spacetime\cite{ks1}\cite{st1}. 
The SGM action (\ref{SGM}) is invariant at least under 
the following symmetry\cite{st2}; 
ordinary GL(4R), 
the following new NLSUSY transformation; 
\begin{equation}
\delta^{NL} \psi(x) ={1 \over \kappa^{2}} \zeta + 
i \kappa^{2} (\bar{\zeta}{\gamma}^{\rho}\psi(x)) \partial_{\rho}\psi(x),
\quad
\delta^{NL} {e^{a}}_{\mu}(x) = i \kappa^{2} (\bar{\zeta}{\gamma}^{\rho}\psi(x))\partial_{[\rho} {e^{a}}_{\mu]}(x),
\label{newsusy}
\end{equation} 
where $\zeta$ is a constant spinor 
and $\partial_{[\rho} {e^{a}}_{\mu]}(x) = 
\partial_{\rho}{e^{a}}_{\mu}-\partial_{\mu}{e^{a}}_{\rho}$, \\
the following GL(4R) transformations due to (\ref{newsusy}); 
\begin{equation}
\delta_{\zeta} {w^{a}}_{\mu} = \xi^{\nu} \partial_{\nu}{w^{a}}_{\mu} 
+ \partial_{\mu} \xi^{\nu} {w^{a}}_{\nu}, 
\quad
\delta_{\zeta} s_{\mu\nu} = \xi^{\kappa} \partial_{\kappa}s_{\mu\nu} 
+ \partial_{\mu} \xi^{\kappa} s_{\kappa\nu} 
+ \partial_{\nu} \xi^{\kappa} s_{\mu\kappa}, 
\label{newgl4r}
\end{equation}
where $\xi^\rho = i \kappa^2 \bar\zeta \gamma^\rho \psi(x)$ 
and $s_{\mu \nu} = w{^a}_\mu w_{a \nu}$, \\
and the following local Lorentz transformation on $w{^a}_{\mu}$; 
\begin{equation}
\delta_L w{^a}_{\mu}
= \epsilon{^a}_b w{^b}_{\mu}
\label{Lrw}
\end{equation}
with the local parameter
$\epsilon_{ab} = (1/2) \epsilon_{[ab]}(x)$    
or equivalently on $\psi$ and $e{^a}_{\mu}$
\begin{equation}
\delta_L \psi(x) = - {i \over 2} \epsilon_{ab}
      \sigma^{ab} \psi,     \quad
\delta_L {e^{a}}_{\mu}(x) = \epsilon{^a}_b e{^b}_{\mu}
      + {\kappa^{4} \over 4} \varepsilon^{abcd}
      \bar{\psi} \gamma_5 \gamma_d \psi
      (\partial_{\mu} \epsilon_{bc}).
\label{newlorentz}
\end{equation}
The local Lorentz transformation forms a closed algebra, for example, 
on $e{^a}_{\mu}(x)$ 
\begin{equation}
[\delta_{L_{1}}, \delta_{L_{2}}] e{^a}_{\mu}
= \beta{^a}_b e{^b}_{\mu}
+ {\kappa^{4} \over 4} \varepsilon^{abcd} \bar{\psi}
\gamma_5 \gamma_d \psi
(\partial_{\mu} \beta_{bc}),
\label{comLr1/2}
\end{equation}
where $\beta_{ab}=-\beta_{ba}$ is defined by
$\beta_{ab} = \epsilon_{2ac}\epsilon{_1}{^c}_{b} 
- \epsilon_{2bc}\epsilon{_1}{^c}_{a}$. 
The commutators of two new NLSUSY transformations (\ref{newsusy}) 
on $\psi(x)$ and ${e^{a}}_{\mu}(x)$ 
are GL(4R), i.e. new NLSUSY (\ref{newsusy}) is the square-root of GL(4R); 
\begin{equation}
[\delta_{\zeta_1}, \delta_{\zeta_2}] \psi
= \Xi^{\mu} \partial_{\mu} \psi,
\quad
[\delta_{\zeta_1}, \delta_{\zeta_2}] e{^a}_{\mu}
= \Xi^{\rho} \partial_{\rho} e{^a}_{\mu}
+ e{^a}_{\rho} \partial_{\mu} \Xi^{\rho},
\label{com1/2-e}
\end{equation}
where 
$\Xi^{\mu} = 2i (\bar{\zeta}_2 \gamma^{\mu} \zeta_1)
      - \xi_1^{\rho} \xi_2^{\sigma} e{_a}^{\mu}
      (\partial_{[\rho} e{^a}_{\sigma]})$.
They show the closure of the algebra. 
SGM action (\ref{SGM}) is invariant at least under\cite{st2}
\begin{equation}
[{\rm global\ NLSUSY}] \otimes [{\rm local\ GL(4,R)}] 
\otimes [{\rm local\ Lorentz}],  \\
\label{sgmsymm}
\end{equation}
which is isomorphic to SP whose single irreducible representation gives 
the group theoretical description of SGM\cite{ks2}.  \par

Here we just mention that the SGM action of Eq.(\ref{SGM}) 
is a nontrivial generalization of E-H action. 
Interestingly the following local spinor translation 
with a local parameter $\epsilon(x)$, 
$\delta \psi = \epsilon$, 
$\delta e{^a}{_\mu} = - i \kappa^4 
(\bar\epsilon \gamma^a \partial_\mu \psi 
+ \bar\psi \gamma^a \partial_\mu \epsilon)$, 
gives $\delta w{^a}_\mu = 0 = \delta w{_a}^\mu$. 
However, this local spinor transformation 
cannot transform away the d.o.f. of $\psi$. 
Indeed, $\psi$ seems to be transformed away 
if we choose $\delta \psi = \epsilon = - \psi$, 
but it is restored precisely in the unified vierbein $w{^a}_\mu$ 
by simultaneously transforming $e{^a}_\mu$, 
i.e., $w(e, \psi) = w(e + \delta e, \psi + \delta \psi) 
= w(e + t, 0)$ as indicated by $\delta w{^a}_\mu = 0$. 
And also the above local spinor transformation is a fake gauge 
transformation in a sense that, in contrast with the local Lorentz 
transformation on the coordinates in the vierbein formalism of EGRT, 
it cannot be eliminate the d.o.f. of $\psi$ 
since the unified vierbein $w{^a}_\mu = e{^a}_\mu + t{^a}_\mu$ 
is the only gauge field on SGM spacetime and contains 
only integer spin. 
This confusive situation comes from the new geometrical formulation 
of SGM on unfamiliar SGM spacetime, where besides the Minkowski 
coordinates $x^a$, $\psi$ is a Grassmann coordinate 
(i.e. the fundamental d.o.f.) defining the tangential spacetime 
with SO(3,1) $\times$ SL(2,C) d.o.f. inspired by NLSUSY. 
The local spinor transformation ($\delta \psi = \epsilon(x)$) 
is just a coordinate transformation(redifinition) on SGM spacetime. 
These situation can be understood easily by observing 
that the unified vierbein $w{^a}_\mu = e{^a}_\mu + t{^a}_\mu$ 
is defined by $\omega^a = d x^a + i \kappa^4 \bar\psi \gamma^a d \psi 
= w^a{}_\mu d x^\mu$, where $\omega^a$ is the NLSUSY invarinat 
differential one-form of V-A\cite{va} 
and $(x^a, \psi)$ are coordinates 
specifying the (SGM) flat spacetime inspired by NLSUSY. \\
From these geometrical viewpoints (in SGM spacetime) 
we can understand that $\psi$ is a coordinate and would 
not be transformed away and the initial SGM spacetime 
is preserved. 
Eliminating $\psi$ by some arguments regarding 
the above local spinor translation as a gauge transformation 
leads to a different theory (ordinary E-H action) 
with a different (Minkowski) flat spacetime, 
which is another from SGM scenario considering that the SGM spacetime 
is an ultimate physical entity.

The linearization of such a high nonlinear theory is interesting and 
inevitable to obtain a renormalizable field theory which is equivalent  
and describes the observed low energy physics.   \\
The flat space linearization of N=1 V-A model has been carried out and proved that N=1 V-A model 
is equivalent to N=1 scalar supermultiplet\cite{ikruz} ${\it or}$ N=1 axial vector gauge supermultiplet 
of linear SUSY\cite{stt2}. \\
As a flat space exercise for the  extended SGM linerization, we have carried out the linearization of  
N=2 V-A model and  shown that it is equivalent to the spontaneously broken N=2 linear SUSY  
${\it vector \ J^{P}=1^{-}}$ gauge supermultiplet model with  SU(2) structure\cite{stt1}. 
Interestingly  SU(2) algebraic gauge structure of the electroweak standard model(SM) 
may be explained for the first time provided that the electroweak gauge bosons are  the composite  fields 
of this(SGM) type in the low energy.  \\
In these works the linearization are carried out by using the superfield formalism 
and/or by the heuristic and intuitive arguments 
on the relations between the component fields of L SUSY and NLSUSY. 
%
%
%
%
%
%
In either case it is crucial to discover  the SUSY invariant relations  which connect the supermultiplets 
of L and NL theories and reproduce the SUSY transformations.   \\
In abovementioned cases of the global SUSY in flat spacetime  the SUSY invariant relations are obtained 
straightforwardly, for L and NL supermultiplets are well undestood and the algebraic structures are the same SP.    \par
The situation  is rather different in SGM, for (i) the supermultiplet structure of 
the linearized theory of SGM is unknown except it is expected to be a broken SUSY SUGRA-like theory 
containing graviton and a (massive) spin 3/2 field as  dynamical d.o.f. and 
(ii) the algebraic structure (\ref{sgmsymm}) is changed into  SP.       \\
%
%
%
%
%
%
Therefore  by the heuristic arguments and referring to SUGRA  we discuss for the moment 
the linearization of N=1 SGM.          \par
%
%
%
%
%
%
At first, we assume  faithfully to SGM scenario that; \\
(i) the linearized theory should contain the  spontaneously broken ${\it global}$ (at least) SUSY  \\
(ii) graviton is an elementary field(not composite of superons coresponding to the vacuum of the Clifford algebra) 
in both L and NL theories   \\
(iii) the NLSUSY supermultiplet of SGM ($e{^a}_{\mu}(x)$, $\psi(x)$)  should be connected 
to the composite supermultiplet 
(${\tilde e}{^a}_{\mu}(e(x), \psi(x))$, ${\tilde \lambda}_{\mu}(e(x), \psi(x))$)  
for elementary graviton field and a composite (massive) spin 3/2 field of the SUGRA-like linearized theory. \par
From these assumptions and following the arguments performed in the flat space cases  we require that 
the SUGRA gauge transformation \cite{fvfdz} with the global spinor parameter ${\zeta}$ 
%
%
%
%
%
%
%
should hold for the supermultiplet (${\tilde e}{^a}_{\mu}(e, \psi)$, ${\tilde \lambda_{\mu}(e, \psi)}$)  
of the (SUGRA-like) linearized theory, i.e.,  \\
\begin{equation}
\delta {\tilde e}{^a}_{\mu}(e, \psi)  
      = i\kappa \bar{ \zeta} \gamma^{a} {\tilde \lambda_{\mu}(e, \psi)}, 
\label{sugral-2}
\end{equation} 
\begin{equation}
\delta {\tilde \lambda}_{\mu}(e, \psi)  
      = {2 \over \kappa}D_{\mu}{ \zeta}  
      = -{i \over \kappa}{{\tilde \omega(e, \psi)}_{\mu}}^{ab}\sigma_{ab}{ \zeta}, 
\label{sugral-3/2} 
\end{equation} 
where  $\sigma^{ab} = {i \over 4}[\gamma^a, \gamma^b]$, 
$D_{\mu}=\partial_{\mu}-{i \over 2} {{\omega}_{\mu}}^{ab}(e, \psi)\sigma_{ab}$, ${ \zeta}$ is a 
global spinor parameter and the variations in the left-hand side are induced by NLSUSY  (\ref{newsusy}). \par
We put the following SUSY invariant relations which connect $e^{a}{_\mu}$ to ${\tilde e}{^a}_{\mu}(e, \psi)$;
\begin{equation}
{\tilde e}{^a}_{\mu}(e, \psi) = { e}{^a}_{\mu}(x).     
\label{relation-2}
\end{equation} 
This relation (\ref{relation-2}) is the assumption (ii) and holds simply the metric conditions.
Consequently the following covariant relation  is obtained by substituting  (\ref{relation-2}) 
into (\ref{sugral-2}) and  computing the variations under (\ref{newsusy})\cite{sts}; 
\begin{equation}
{\tilde \lambda}_{\mu}(e, \psi)  
      = \kappa \gamma_{a} \gamma^{\rho} \psi(x) \partial_{[\rho} e{^a}_{\mu]}.    
\label{relation-3/2} 
\end{equation} 
(As discussed later these should may be considered as the leading order of the expansions in $\kappa$ of 
SUSY invariant relations. The expansions terminate with $(\psi)^{4}$. 
Now we see LSUSY transformation 
%
%
induced by (\ref{newsusy}) on the (composite) supermultiplet 
(${\tilde e}{^a}_{\mu}(e, \psi)$, ${\tilde \lambda}_{\mu}(e, \psi$)).    \\
The LSUSY transformation  on $\tilde e{^a}_{\mu}$ becomes  as follows. 
The left-hand side of (\ref{sugral-2}) gives
\begin{equation}
\delta {\tilde  e}{^a}_{\mu}(e, \psi) = \delta^{NL} {e^{a}}_{\mu}(x) 
= i \kappa^{2} (\bar{\zeta}{\gamma}^{\rho}\psi(x))\partial_{[\rho} {e^{a}}_{\mu]}(x). 
\label{susysgm-2} 
\end{equation} 
While substituting (\ref{relation-3/2}) into the right-hand side 
of (\ref{sugral-2}) we obtain 
\begin{equation}
i \kappa^{2} (\bar{\zeta}{\gamma}^{\rho}\psi(x))\partial_{[\rho} {e^{a}}_{\mu]}(x) + \cdots({\rm extra \ terms}).
\label{susysgm-3/2} 
\end{equation} 
These results show that  (\ref{relation-2}) and (\ref{relation-3/2}) are not  SUSY invariant relations 
and reproduce (\ref{sugral-2}) with unwanted extra terms which should 
be identified with the auxiliary fields. 
The commutator of the two LSUSY transformations induces GL(4R) with the field dependent parameters as follows;
\begin{equation}
[\delta_{\zeta_1}, \delta_{\zeta_2}]{\tilde  e}{^a}_{\mu}(e, \psi) 
= \Xi^{\rho} \partial_{\rho} {\tilde  e}{^a}_{\mu}(e, \psi)    
+ {\tilde  e}{^a}_{\rho}(e, \psi)\partial_{\mu} \Xi^{\rho},
\label{susysgmcom-2}
\end{equation}
where 
$\Xi^{\mu} = 2i (\bar{\zeta}_2 \gamma^{\mu} \zeta_1)
      - \xi_1^{\rho} \xi_2^{\sigma} e{_a}^{\mu}
      (\partial_{[\rho} e{^a}_{\sigma]})$.

On  ${\tilde \lambda}_{\mu}(e, \psi)$, the left-hand side of (\ref{sugral-3/2}) becomes 
apparently rather complicated; 
\ba
\delta {\tilde \lambda}_{\mu}(e, \psi)  
\A = \A {\kappa } \delta( \gamma_{a} \gamma^{\rho} \psi(x) \partial_{[\rho} e{^a}_{\mu]}) \nonu
\A = \A {\kappa } \gamma_{a}[ \delta^{NL}\gamma^{\rho} \psi(x) \partial_{[\rho} e{^a}_{\mu]} + 
  \gamma^{\rho} \delta^{NL} \psi(x) \partial_{[\rho} e{^a}_{\mu]} +
  \gamma^{\rho}  \psi(x) \partial_{[\rho} \delta^{NL} e{^a}_{\mu]}]. 
\label{susysgm-3/2} 
\ea
However the commutator of the two LSUSY transformations induces the similar GL(4,R);
\begin{equation}
[\delta_{\zeta_1}, \delta_{\zeta_2}]{\tilde \lambda}_{\mu}(e, \psi)  
= \Xi^{\rho} \partial_{\rho} {\tilde \lambda}_{\mu}(e, \psi)  
+ {\tilde \lambda}_{\rho}(e, \psi)\partial_{\mu} \Xi^{\rho}.  
\label{susysgmcom-3/2}
\end{equation}                  
These results indicate that it is necessary to generalize  (\ref{sugral-2}), (\ref{sugral-3/2}) and 
(\ref{relation-3/2})  for obtaining SUSY invariant relations and for the closure of the algebra.
%
%
%
%
%
%
Furthermore  due to the complicated expression of LSUSY (\ref{susysgm-3/2}) 
which makes the physical and 
mathematical structures are obscure, we can hardly guess a linearized invariant action 
which is equivalent to SGM.       \par
Now we generalize the linearization by considering the auxiliary fields 
such that LSUSY transformation on the linearized fields 
induces SP transformation.   \par
%
%
%
%
%
%
By comparing (\ref{sugral-3/2}) with (\ref{susysgm-3/2}) we understand that the local Lorentz transformation  
plays a crucial role. 
As for the local Lorentz transformation on the linearized asymptotic fields corresponding 
to the observed particles (in the low energy), 
it is natural to take (irrespective of (\ref{newlorentz})) the following forms   \\ 
\begin{equation}
\delta_L \tilde \lambda_{\mu}(x) = - {i \over 2} { \epsilon}_{ab}
      \sigma^{ab} \tilde \lambda_{\mu}(x),     \quad
\delta_L \tilde { e^{a}}_{\mu}(x) = {\epsilon}{^a}_b \tilde e{^b}_{\mu}, 
\label{lorentz}
\end{equation}
where $\epsilon_{ab} = (1/2)\epsilon_{[ab]}(x)$ is a local parameter.   
The relation between (\ref{newlorentz}), i.e. the ultimate Lorentz invariance encoded 
geometrically in  SGM space-time,  and (\ref{lorentz}), i.e.  the  Lorentz invariance 
defined on the (composite) asymptotic field in Riemann space-time, is unclear. 
In SGM  the local Lorentz transformations  (\ref{Lrw}) 
and (\ref{newlorentz}), 
i.e. the local Lorentz invariant gravitational interaction of superon, 
are introduced   by the geomtrical arguments in SGM spacetime\cite{st2} following  EGRT.  
While in SUGRA theory the local Lorentz  transfomation  invariance  (\ref{lorentz}) 
is  realized as usual by introducing  the Lorentz spin connection $\omega{_\mu}^{ab}$ . 
And the LSUSY transformation is defined successfully by the (Lorentz) covariant 
derivative containing the spin connection $\tilde \omega_{\mu}{^{ab}}(e,\psi)$ as seen in (\ref{sugral-3/2}), 
which causes the super-Poincar\'e algebra on the commutator of SUSY and is convenient for 
constructing the invariant action. 
Therefore in the linearized (SUGRA-like) theory the local Lorentz  transformation  invariance is expected 
to be realized as usual by defining (\ref{lorentz}) and introducing the Lorentz spin connection $\omega{_\mu}^{ab}$.
We investigate how the spin connection ${\tilde \omega}{_\mu}^{ab}(e,\psi)$ appears 
in the linearized (SUGRA-like) theory through  the linearization process. 
This is also crucial for constructing a nontrivial (interacting) linearized action 
which has manifest invariances.  \par  
We discuss the Lorentz covariance of the  transformation by comparing (\ref{susysgm-3/2}) with 
the right-hand side of (\ref{sugral-3/2}). 
%
%
%
The direct computation of (\ref{sugral-3/2}) by using SUSY invariant relations (\ref{relation-2}) 
and (\ref{relation-3/2}) under (\ref{newsusy}) produces complicated redundant terms 
as read off from (\ref{susysgm-3/2}).
The local Lorentz invariance of the linearized theory may become ambiguous and lose the manifest invariance. \\
For a simple  restoration of the manifest local Lorentz invariance 
we survey the possibility that such redundant terms may be adjusted  by  the d.o.f of 
the auxiliary fields in the linearized supermultiplet.
As for the auxiliary fields it is necessary for the closure of the off-shell superalgebra 
to include the equal number of the fermionic and the bosonic d.o.f. in the linearized supermultiplet. 
As new NLSUSY is a global symmetry, ${\tilde \lambda}_{\mu}$ has 16 fermionic d.o.f.. 
Therefore at least 4 bosonic d.o.f. must be added to the off-shell SUGRA supermultiplet 
with 12 d.o.f.\cite{swfv} and a vector field  may be a simple candidate.  \par
However, counting the bosonic d.o.f. present in the redundant terms corresponding to 
${\tilde \omega}{_\mu}^{ab}(e,\psi)$, 
we may need a bigger supermultiplet  e.g. $16 + 4 \cdot 16 = 80$  d.o.f., to carry out the linearization, 
in which case a rank-3 tensor $\phi_{\mu\nu\rho}$ 
and a rank-2 tensor-spinor $\lambda_{\mu\nu}$ may be candidates for the auxiliary fields.     \par
Now we consider  the  simple modification of SUGRA transformations(algebra) by adjusting 
the (composite) structure of the (auxiliary) fields.  
We take, in stead of (\ref{sugral-2}) and  (\ref{sugral-3/2}), 
\begin{equation}
\delta {\tilde e}{^a}_{\mu}(x) 
      = i\kappa \bar{\zeta} \gamma^{a} {{\tilde \lambda}_{\mu}(x)} + \bar{\zeta}{\tilde \Lambda}{^a}_{\mu}, 
\label{newsugral-2}
\end{equation} 
\begin{equation}
\delta {\tilde \lambda}_{\mu}(x) 
      = {2 \over \kappa}D_{\mu}\zeta + {\tilde \Phi}_{\mu}\zeta  
      = -{i \over \kappa}{\tilde \omega}{_{\mu}}^{ab}\sigma_{ab}\zeta + {\tilde \Phi}_{\mu}\zeta , 
\label{newsugral-3/2} 
\end{equation} 
where  $\tilde \Lambda{^a}_{\mu}$ and  $\tilde \Phi_{\mu}$ represent symbolically the auxiliary fields 
80+80 and are functionals of $e^{a}{_\mu}$ and $\psi$. 
%
%
%
%
%
%
We need $\tilde \Lambda{^a}_{\mu}$ term in (\ref{newsugral-2}) to alter (\ref{susysgm-2}), 
(\ref{susysgmcom-2}), (\ref{susysgm-3/2}) and  (\ref{susysgmcom-3/2}) 
toward that of super-Poincar\'e algebra of SUGRA.  
%
%
%
%
%
%
We attempt the restoration of the manifest local Lorentz invariance order by order by adjusting 
$\tilde \Lambda{^a}_{\mu}$ and  $\tilde \Phi_{\mu}$. 
%
%
%
%
%
%
In fact, the Lorentz spin connection  ${\omega}{_\mu}^{ab}(e)$(i.e. the leading order terms of 
${\tilde \omega}{_\mu}^{ab}(e, \psi)$) of (\ref{newsugral-3/2}) is reproduced by taking the following one  
%
%
%
%
%
%
\begin{equation}
\tilde \Lambda{^a}_{\mu} = {\kappa^{2} \over 4}[ ie_{b}{^\rho}\partial_{[\rho}e{^b}_{\mu]}\gamma{^a}\psi 
- \partial_{[\rho}e_{\mid b \mid \sigma]}e{^b}_{\mu}\gamma^{a}\sigma^{\rho\sigma}\psi ], 
\label{auxlambda-1}
\end{equation} 
which holds (\ref{susysgmcom-2}).
Accordingly $\tilde \lambda_{\mu}(e,\psi)$ 
%
%
%
%
%
%
%
%
%
is determined up to the first order in $\psi$ as follows; 
\begin{equation}
\tilde \lambda_{\mu}(e,\psi) 
= { 1 \over 2i\kappa}( i\kappa^{2}\gamma_{a} \gamma^{\rho} \psi(x) \partial_{[\rho} e{^a}_{\mu]}
- \gamma_{a}\tilde \Lambda{^a}_{\mu} ) = -{i \kappa \over 2}{{\omega}_{\mu}}^{ab}(e)\sigma_{ab}\psi,  
\label{lambda-o1}
\end{equation} 
which indicates the minimal Lorentz covariant gravitational interaction of superon. 
Sustituting  (\ref{lambda-o1}) into (\ref{newsugral-3/2}) we obtain the following new LSUSY transformation 
of $\tilde \lambda_{\mu}$(after Fiertz transformations) 
\ba
\delta {\tilde \lambda}_{\mu}(e,\psi) 
\A = \A -{i \kappa \over 2} \{ \delta^{NL}{{\omega}_{\mu}}^{ab}(e)\sigma_{ab}\psi + 
{{\omega}_{\mu}}^{ab}(e)\sigma_{ab} \delta^{NL}\psi \}  \nonu
\A = \A -{i \over {2 \kappa}} {{\omega}_{\mu}}^{ab}(e)\sigma_{ab} \zeta + 
{i \kappa \over 2} \{ \tilde \epsilon^{ab}(e,\psi)\sigma_{ab}\cdot{}\omega_{\mu}{^{cd}}(e)\sigma_{cd}\psi + \cdots \}.\label{varlambda-o1} 
\ea 
Remarkably the local Lorentz transformations  of  ${\tilde \lambda}_{\mu}(e,\psi)$ (,i.e. the second term)
with the field dependent antisymmetric parameters $\tilde \epsilon^{ab}(e, \psi)$ is induced 
in addition to the intended ordinary global SUSY transformation. 
(\ref{lambda-o1}) is the SUSY invariant relations for $\tilde \lambda_{\mu}(e,\psi)$ in the lowest 
order with $\psi$, 
for the SUSY transformation of (\ref{lambda-o1}) gives the right hand side of (\ref{newsugral-3/2}) 
with the consequent auxiliary fields. 
Interestingly  the commutator of the two LSUSY transformations  on (\ref{lambda-o1}) induces 
GL(4R); 
\begin{equation}
[\delta_{\zeta_1}, \delta_{\zeta_2}]{\tilde \lambda}_{\mu}(e, \psi)  
= \Xi^{\rho} \partial_{\rho} {\tilde \lambda}_{\mu}(e, \psi)  
+ \partial_{\mu} \Xi^{\rho}{\tilde \lambda}_{\rho}(e, \psi),  
\label{varlambda-o1-comm}
\end{equation}                   
where $\Xi^{\rho}$ is the same field dependent parameter as given in (\ref{susysgmcom-2}).
(\ref{susysgmcom-2}) and (\ref{varlambda-o1-comm}) show the closure of the algebra 
on SP algebra provided that the SUSY invariant relations (\ref{relation-2}) and (\ref{lambda-o1}) 
are adopted. 
These phenomena coincide with SGM scenario\cite{ks1}\cite{ks2} from the algebraic point of view, 
i.e. they are the superon-graviton composite (eigenstates) corresponding to the  linear representations 
of SP algebra. 
As for the redundant terms in (\ref{varlambda-o1}) with $(\psi)^{2}$ 
%
%
%
the SUSY transformations 
%
%
%
we can recast  them by considering the modified spin connection ${\tilde \omega}{_\mu}^{ab}(e, \psi)$ 
particularly with the contorsion terms and (the auxiliary field) $\tilde \Phi_{\mu}(e,\psi)$. 
In fact, we have proved that the contributions to (\ref{varlambda-o1}) 
from the SUGRA-inspired contorsion terms; 
\ba 
K_{\mu ab}\sigma^{ab}\psi 
\A = \A {i \kappa^{2} \over 4}(\bar{\tilde \lambda}_{a}\gamma_{b}\tilde \lambda_{\mu}-
\bar{\tilde \lambda}_{b}\gamma_{a}\tilde \lambda_{\mu} + 
\bar{\tilde \lambda}_{a}\gamma_{\mu}\tilde \lambda_{b})\sigma^{ab}\psi   \nonu
\A = \A {i\kappa^{4} \over 16} 
\{ (e_{a}{^\mu}\bar\psi\sigma^{cd}\omega_{\nu cd}(e)\gamma_{b}\omega_{\mu fg}(e)\sigma^{fg}\psi-
[a \leftrightarrow b] + \cdots \}\sigma^{ab}\psi, 
\label{contorsion}
\ea
satisfies 
\begin{equation}
[\delta_{\zeta_1}, \delta_{\zeta_2}] (K_{ab\mu}\sigma^{ab}\psi)  
=  \Xi^{\rho} \partial_{\rho}(K_{ab\mu}\sigma^{ab}\psi)   
+ \partial_{\mu} \Xi^{\rho}(K_{ab\rho}\sigma^{ab}\psi).  
\label{contorsion-comm}
\end{equation}
We can obtain the complete linearized (off-shell) supermultiplets of the super-Poincar\'e algebra 
by repeating the similar procedures (including  the auxiliary fields ${80+80}$) order by order 
which terminates with $(\psi)^{4}$. 
The complicated procedures has been carried out successfully up to $O(\psi^{2})$\cite{sts2}.  \par
%
%
%
%
%
Finally we mention  another way of the systematic linearization by using the superfield formalism 
applied to the coupled system of V-A action with SUGRA\cite{lr}. We can define on such a coupled system 
a local spinor gauge symmetry which induces a super-Higgs mechanism\cite{dz} 
converting V-A field to the longitudinal component of massive spin 3/2 field. 
The consequent Lagrangian obtained may be an analogue that we anticipate 
in above discussions but with the elementary spin 3/2 field. 
%
%
%
%
%
%
\\
The linearization of the new E-H type action (\ref{SGM}) 
with the extra dimensions, 
which gives another unification framework describing the observed particles as elementary fields, is open. 
And the linearization of SGM action for spin 3/2 N-G fermion field\cite{st3} (with extra dimensions) 
may be in the same scope.  \par
Now we summarize the results as follows. 
Referring to SUGRA transformations we have attempted explicitly the linearization of N=1 SGM 
up to $O(\psi^2)$ in the (SUGRA-like) LSUSY transformations. 
The closure of the new LSUSY transformations  (\ref{newsugral-2}) and  (\ref{varlambda-o1}) 
on the linearized supermultiplet, which are different from SUGRA transformations, 
can be proved order by order with $\psi$ by introducing the auxiliary fields.
It is interesting that the simple relation $\lambda_{\mu}=e^{a}{_\mu}\gamma_{a}\psi + \cdots $, 
which is sugested by the flat spacetime linearization, seems disfavour with the SGM linearization. 
As conjectured before, what LSUSY SP may be to SGM in quantum field theory, 
what O(4) symmetry is to the relativistic hydrogen model in quantum mechanics, 
which is tested  by the linearization. 
The linearization of SGM is physically interesting in general, 
even if it were  a existing SUGRA-like theory, 
for the consequent broken LSUSY theory 
is shown to be equivalent 
and gives a new insight into the fundamental structure of 
nature (like the relation between  Landau-Ginzberg theory 
 and  BCS theory for supercnductivity).
The complete linearization to all orders, 
which can be anticipated by the systematics emerging in the present study, 
needs specifications of the auxiliary fields 
and remains to be studied. They are now in progress.

\vskip 10mm

The authors  would like to thank U. Lindstr\"om for the interest in our works and for bringing the reference 
to our attentions and Y. Tanii for useful discussions. 
The work of M. Sawaguchi is supported in part by the research project of  High-Tech Research Center 
of Saitama Institute of Technology.

\newpage

%
\newcommand{\NP}[1]{{\it Nucl.\ Phys.\ }{\bf #1}}
\newcommand{\PL}[1]{{\it Phys.\ Lett.\ }{\bf #1}}
\newcommand{\CMP}[1]{{\it Commun.\ Math.\ Phys.\ }{\bf #1}}
\newcommand{\MPL}[1]{{\it Mod.\ Phys.\ Lett.\ }{\bf #1}}
\newcommand{\IJMP}[1]{{\it Int.\ J. Mod.\ Phys.\ }{\bf #1}}
\newcommand{\PR}[1]{{\it Phys.\ Rev.\ }{\bf #1}}
\newcommand{\PRL}[1]{{\it Phys.\ Rev.\ Lett.\ }{\bf #1}}
\newcommand{\PTP}[1]{{\it Prog.\ Theor.\ Phys.\ }{\bf #1}}
\newcommand{\PTPS}[1]{{\it Prog.\ Theor.\ Phys.\ Suppl.\ }{\bf #1}}
\newcommand{\AP}[1]{{\it Ann.\ Phys.\ }{\bf #1}}

\end{document}